# Measurement Errors as Bad Leverage Points

Eric Blankmeyer

Texas State University
eb01@txstate.edu

March 2020

**Abstract.** Errors-in-variables is a long-standing, difficult issue in linear regression; and progress depends in part on new identifying assumptions. I characterize measurement error as bad-leverage points and assume that fewer than half the sample observations are heavily contaminated, in which case a high-breakdown robust estimator may be able to isolate and downweight or discard the problematic data. In simulations of simple and multiple regression where eiv affects 25% of the data and $R^2$ is mediocre, certain high-breakdown estimators have small bias and reliable confidence intervals. A brief discussion of heteroscedasticity and robust estimation is provided in an appendix.

# Measurement Errors as Bad Leverage Points

## 1. Introduction

A venerable issue in linear-regression analysis is errors in variables (eiv, also called measurement error), when a regressor is not directly observable. Instead, a proxy is available that differs from the regressor because of random contamination. In ordinary least squares (ols) estimation, eiv produces a bias that does not vanish asymptotically. Many researchers attest that this is a pervasive and challenging problem. Theil (1971, p. 607-613) reviews some procedures for dealing with eiv, "none of which is really simple in application." According to Malinvaud (1980, p. 416), "If in many cases structural parameters can be identified, consistent estimators appropriate to such cases are virtually useless in econometrics, since there are too few data." Friedman (1992, p. 2131) comments that "the common practice is to regress a variable Y on a vector of variables X and then accept the regression coefficients as supposedly unbiased estimates of structural parameters, without recognizing that all variables are only proxies for the variables of real interest, if only because of measurement error, though generally also because of transitory factors that are peripheral to the subject under consideration. I suspect that the regression fallacy is the most common fallacy in the statistical analysis of economic data, alleviated only occasionally by consideration of the bias introduced when 'all variables are subject to error.' " Dagenais and Dagenais (1997, p. 195) note that, for ols, "intended 95% confidence intervals may in practice turn out to be almost 0% intervals, even when the errors of measurement are not exceedingly large….Similarly, Student t-tests using the critical values corresponding normally to 5% type I errors may in fact correspond to tests with type I errors of size equal to almost 100%! This may have dramatic consequences since one may be induced to reject a null hypothesis when this hypothesis is true, with a probability close to 100%!" Greene (2003, p. 84) remarks that the "general assessment of the problem is not particularly optimistic. The biases introduced by measurement error can be rather severe." And Hausman (2001, p. 58) speaks of " the 'Iron Law of Econometrics' –the magnitude of the estimate is usually smaller than expected."

Eiv models assume that *all the observations* are potentially contaminated by measurement error in one or several regressors. Since in many situations this assumption will be unduly pessimistic, I explore eiv estimation when egregious measurement errors affect *only a minority of observations* –a subset that can be characterized as bad-leverage points. A robust high-breakdown estimator can then locate and downweight these bad-leverage observations. Most of my simulations emulate data that are noisy due to eiv and also because $R^2$ for the correctly-measured variables is not very high. In this challenging framework, which may be fairly typical of cross-section data, the minimum covariance determinant procedure seems to perform adequately whereas a robust estimator designed specifically for linear regression is less successful in terms of bias control and accurate coverage of a confidence interval.

Eiv has generated a vast literature, and a comprehensive review is beyond this essay's scope. Recent research has focused on the treatment of measurement error in nonlinear and nonparametric models as surveyed by Carroll et al. 2006, Chen et al. 2011, and Schennach 2004a, 2004b. High-breakdown estimation could be considered for some of these models, but significant conceptual difficulties remain (Stromberg and Ruppert 1992, Stromberg 1993, Davies and Gather 2005). This paper is therefore limited to the linear eiv model, with which several surveys are wholly or partly concerned, e. g., Fuller 1987, Cheng and Ness 1999, Wansbeek and Meijer 2000, and Schennach 2016.

Section 2 is a concise review of the classical linear eiv model and several of the estimators proposed for it. In Section 3 measurement errors are interpreted as bad-leverage observations; and, subject to an identifying assumption, high-breakdown estimators are proposed to cope with eiv. Six simulations are presented and discussed in Section 4, followed by a brief examination of a real data set in Section 5. Section 6 contains a few caveats and conclusions. In an appendix I explore the relationship between heteroscedasticity and outlying observations.

## 2. A canonical eiv model

Although researchers have explored many variations of the measurement-error problem in linear regression, I begin with a canonical bivariate eiv model:
$$y_i = \alpha + \beta x_i + u_{yi} \quad , \tag{1}$$
where $\alpha$ and $\beta$ are unknown parameters, $y_i$ is an observation on the dependent variable, and $u_{yi}$ is an unobservable random variable, independently and identically distributed (iid) with zero expectation and standard deviation $\sigma_y$. The regressor $x_i$ is also unobservable; instead a researcher observes $x_i + u_{xi}$, where $u_{xi}$ is an iid error with zero expectation and standard deviation $\sigma_x$. It is assumed that $x_i$, $u_{xi}$ and $u_{yi}$ are stochastically independent of one another. The object is to obtain consistent estimates of $\alpha$, $\beta$, and $\sigma_y$ from a random sample of $y_i$ and $x_i + u_{xi}$. The eiv problem is that $x_i$ and $u_{xi}$ are never observed separately but always as $x_i + u_{xi}$, a mismeasured regressor that is correlated with the regression disturbance $u_{yi} - \beta u_{xi}$. Consequently, the OLS estimate of $\beta$ is inconsistent: it converges to $\beta/(1 + \sigma_x^2 / \text{plim}(\Sigma x_i^2/n))$, so the bias is toward 0 –the notorious "least-squares attenuation." In multiple linear regression eiv bias in one regressor may skew all the estimated coefficients; in addition, it can happen that several regressors are mismeasured. In either case, the direction of ols biases becomes problematic in general (Greene 2003, p. 85-86).

Moreover, if $x_i$ and $u_{xi}$ are both Gaussian, the parameters of interest are not identifiable, hence the non-uniqueness of the maximum-likelihood estimator in the absence of additional information or assumptions. In model (1), identification is achieved if one knows (or is willing to guess) the value of $\sigma_x^2$ or $\text{plim}(\Sigma x_i^2/n)$ or their ratio. For example, data from a validation sample or a replication might yield information about the size of the measurement error. In particular, if the researcher believes that $\sigma_x^2 / \sigma_y^2 \sim 1$, then the maximum-likelihood estimator is the orthogonal regression, the eigenvector corresponding to the smallest eigenvalue of the covariance matrix of $y_i$ and $x_i + u_{xi}$ (but cp. Carroll and Ruppert 1996). Latent-variable models and factor analysis have also been used extensively to model measurement error. In that methodology, identification is of course dependent on judgments about the number of factors to be included and the choice of a "rotation" criterion.

The eiv literature explores several other strategies for the identification and consistent estimation of linear models, some of which can be interpreted as instrumental variables and therefore reflect the strengths and weaknesses of that procedure. For example, it is suggested that in model (1) the sample data be split into groups according to some *a priori* criterion (the instrument); then the regression be performed on the group means in the belief that the $u_{xi}$ will average to zero within each group, at least asymptotically. As Malinvaud (1980, p. 416-419) explains, consistent estimation makes "two demands which are often contradictory. For it is necessary that" the groups be chosen independently of the $u_{xi}$ but also in a way that the group means of the dependent variable do not all converge to $E(y_i)$ since there would then be little or no variation in the dependent variable. As an alternative to grouping, it is suggested that the instrument be formed from the ranking of $x_i + u_{xi}$ in the hope that the ranks will be independent of the $u_{xi}$ but strongly correlated with the $x_i$.

Another instrumental-variable approach achieves identification by assuming that, while the $u_{xi}$ are normally distributed, the $x_i$ are not. Then instruments can be generated from the higher-order moments of the observations $x_i + u_{xi}$, e. g., skewness and kurtosis. Important contributions to this literature include Dagenais and Dagenais 1997 and Erickson and Whited 2002. Other strategies to obtain point estimates or useful bounds on the parameters in the linear eiv model have been proposed in papers by several authors including Frisch 1929, Klepper and Leamer 1984, Black et al. 2000, Cheng and Riu 2006 and Lewbel 2012.

Although one cannot expect a universally acceptable strategy for identification in situations involving measurement error, progress on the eiv problem depends on the formulation of new identifying assumptions that are credible for a well-defined but reasonably wide range of real data sets. In the sequel I offer an identifying strategy that has not, as far as I know, been proposed before.

## 3. Bad leverage points and high-breakdown estimators

As previously mentioned, OLS is a biased estimator of model (1) because $x_i$ and $u_{xi}$ are never observed separately; all the observations are potentially contaminated. While this premise is no doubt realistic in some contexts, it seems unduly pessimistic for many actual data sets, where egregious measurement error may well be limited to a minority of the observations. The robust high-breakdown estimators used in the sequel can in principle cope with contamination in as much as 50% of the data. The rationale for this upper bound is that, when it comes to avoiding very large biases ("breakdown"), no affine-equivariant estimator for linear regression can distinguish between valid and invalid observations if the latter are in the majority (Rousseeuw and Leroy 1987, chapters 1 and 3; Maronna et al. 2006, chapters 3, 5 and 6). Accordingly, *the new identifying assumption is that eiv affects less than half of the sample; in a majority of observations, $u_{xi}$ is negligible. The estimation strategy is simply to use high-breakdown methods that can detect and downweight or eliminate the mismeasured observations, those for which $u_{xi}$ is not negligible.*

Now Rousseeuw and Van Driessen (2006, p. 29) offer a taxonomy of outliers: a point for which $y_i$ diverges from the linear pattern of the majority of the data but whose regressors are not outlying is called a *vertical outlier*. A point with one or more outlying regressors is a *leverage point*. A *good* leverage point lies far from the majority of observations but near to the regression plane implied by the majority. A *bad* leverage point lies far from the majority of observations and their implied regression plane. "Summarizing, a data set can contain four types of points: regular observations, vertical outliers, good leverage points, and bad leverage points. Of course, most data sets do not have all four types."

*In model (1), a bad-leverage observation occurs when variation in a regressor is not matched by a corresponding variation in the dependent variable. Measurement error produces bad-leverage points because $u_{xi}$ is uncorrelated with $y_i$.* Figure 1 displays a pseudo sample of 2000 bivariate observations, 25% of which is contaminated with eiv. The correctly-measured data are concentrated in the central ellipse whose principal axis has a slope of 1 approximately. The mismeasured observations mostly protrude horizontally to the left and right of the central ellipse; they are the bad-leverage points -- the $u_{xi}$ -- whose excess variation flattens out or attenuates the ols slope estimate. Accordingly, I interpret eiv as a type of bad-leverage observation. If the proportion of mismeasured observations is not excessive, an appropriate high-breakdown estimator will focus on the data clustered in the central ellipse of Figure 1 and will therefore produce a good estimate of the regression line.

There is some precedent for my identification strategy. Time-series analysts routinely discover in their data a few "break points" or "level shifts" which they may handle by inserting dummy-variable regressors. The robustness literature has long recognized these anomalies as bad leverage points and has proposed algorithms for detecting them when, as is typically the case, *they are a small proportion of the sample observations* (Rousseeuw and Leroy 1987, 273-284; Maronna et al. 2006, chapter 8; Muler et al. 2009; Kaluzny 2017). It is also noteworthy that the eiv literature discusses consistent estimation based on a validation sample, "a subsample of the original sample for which accurate measurements are available" (Hu and Ridder 2012, p. 348). In my identification strategy, the validation sample is just the uncontaminated majority of the observations, unknown *a priori* but potentially identifiable by a high-breakdown estimator.

The high-breakdown estimators in my simulations are listed in Table 1 together with the sources that describe their algorithms, properties and implementations in the R programming language. MM and DetMCD produce an initial ("raw") estimate that is very robust and has rather low statistical efficiency; i.e., this estimate often downweights some regular observations and good-leverage points along with the vertical outliers and bad-leverage points. The algorithms then perform iterations intended to reinstate the valid data and thereby boost the efficiency of the final estimate. However, my simulations focus on the initial estimates. This is because all the simulations except Table 2 are designed to generate challenging and realistic samples in which *the "true" linear relationship between $x_i$ and $y_i$ is mediocre*, with R-squared in the range of 0.30-0.35. Cross-section data in economics and other fields are frequently quite noisy ($\sigma_y$ is relatively large), and preliminary work indicated that the efficient high-breakdown estimators tend to retain unacceptably large eiv biases in these difficult environments. After all, a high breakdown point guarantees that the estimator's bias is finite but not that it is small. Nowadays, moreover, data sets often have a great many observations, in which case an estimator's efficiency is less important than its ability to control bias.

Table 1 includes only three of the many estimators that have been proposed for robust linear regression; several others are discussed by Rousseeuw and Leroy 1987, Rousseeuw and Croux 1993, Rousseeuw and Hubert 1999, Wang and Raftery 2002, Atkinson et al. 2004, Maronna et al. 2006, Olive and Hawkins 2008, and Park et al. 2012.

This paper is not the first to juxtapose eiv and high-breakdown estimation. Previous work includes Rousseeuw and Leroy 1987, p. 284-285; Zamar 1989; Fekri and Ruiz-Gazen 2004, 2006; Maronna 2005; and Jung 2007. In general, however, these authors consider bad-leverage points to be distinct from measurement error, whereas I see no difference in practice. Adopting the canonical assumption that measurement error affects the entire sample, the authors propose high-breakdown algorithms for orthogonal regression to deal with a limited number of vertical outliers and bad-leverage points. On the other hand, I assume that eiv seriously impacts less than half the sample, where it appears as bad-leverage points. I therefore dispense with the additional assumptions required by orthogonal regression.

## 4. Simulations

This section reports six simulations that are variations of the canonical eiv model. Each simulation has these characteristics: 0 is the value of the intercept α; 1 is the value of the true slope coefficient(s) β; and the sample is replicated 1000 times. For each sample the regressor(s) are contaminated with measurement error in 25 percent of the observations. MM is a linear regression algorithm and therefore generates the initial (or raw) coefficients directly. DetMCD produces an initial covariance matrix **C** from which I compute regression coefficients using the OLS normal equations; thus for equation (1), the DetMCD slope estimate is $c_{xy} / c_{xx}$. Likewise the DetS slope is calculated from a robust covariance matrix which can be considered an initial estimate since S estimators have a high breakdown point but relatively low efficiency.

For the slope coefficient(s), tables 2 through 7 display the bias, the root mean squared error (rmse), and a two-sided 95 percent confidence interval. As for notation, in the text and tables $z \sim N(\mu,\sigma)$ denotes a normally-distributed random variable z with expectation μ and standard deviation σ; and n denotes the sample size.

Table 2 reports a simulation of the bivariate eiv model (1) in which the correlation between $y_i$ and $x_i$ is rather high: $R^2 = 0.80$; specifically, $u_{yi} \sim N(0,1)$ and $x_i \sim N(0,2)$. In addition, $u_{xi} \sim N(10,4)$ --the expected value of the measurement error is not zero, which Malinvaud (1980, pp. 384-385) proposes as a plausible departure from the canonical model. For 200 observations, OLS has negligible sampling error but a large downward bias (attenuation) and a confidence interval with no coverage of the true slope coefficient. MM, DetMCD and DetS are essentially unbiased, and the confidence intervals of all three high-breakdown estimators cover the true slope coefficient.

When n = 2000 in Table 2, the OLS results are essentially unchanged. The three high-breakdown estimators are practically unbiased; and although their confidence intervals are shorter in this larger sample, they still contain the true slope coefficient –although barely so for MM and DetS.

For the bivariate regression in Table 3, the correlation between $y_i$ and $x_i$ is mediocre: $R^2 = 0.310$, which may be more typical of cross-section data sets. Again OLS has a large bias and no coverage, but now MM performs no better than OLS at both sample sizes. When n = 200, DetS has the smallest bias and rmse and also the shortest confidence interval that covers the true slope coefficient; but when n = 2000, DetMCD might be preferred since its confidence interval is more nearly centered on β.

Table 4 summarizes a multiple linear regression for which $R^2$ would be about 0.35 in the absence of eiv. The regressor z is free of measurement error and has a correlation of 0.40 with the regressor x, but $u_{xi}$ contaminates 25 percent of the x data. DetMCD has the smallest bias and is the only estimator both of whose confidence intervals include the true slope coefficients.

The simulation for Table 5 is identical to Table 4 except that the regressor z is also mismeasured: eiv affects each regressor in 12.5 percent of its observations, and the two contaminated sets do not overlap. DetMCD has the smallest biases, but DetS might be preferred since its biases are not much larger while its rmse's are smaller and its confidence intervals are accordingly shorter.

Under the canonical eiv assumption that measurement error potentially contaminates all the observations, a high-breakdown estimator is likely to be biased because it cannot cope with so many bad leverage points. On the other hand, Tables 6 and 7 explore what may happen when procedures often advocated for eiv are applied to data sets in which measurement error affects only 25 percent of the sample. In Table 6 the errors in x and y have the same standard deviation so orthogonal regression (OR) would be appropriate if all the observations were subject to measurement error. The geometric mean regression (GEOM), which estimates the slope as the standard deviation of y divided by the standard deviation of x with the correlation determining the sign, also has a long history in the eiv literature (Frisch 1929, Samuelson 1942, Tofallis 2008). Table 6 shows that the OLS slope estimate is attenuated as usual while OR and GEOM overestimate the slope. Det MCD has the smallest bias and rmse, and its confidence interval contains β.

In Table 7 x is a chi-square variable with 4 degrees of freedom while $u_{xi}$ is Gaussian. Dagenais and Dagenais (1997) propose to use the skewness in x as an instrumental variable (IV), an effective strategy except that $u_{xi}$ again impacts only 25 percent of the sample. The IV estimator performs better than OLS, but DetMCD still has the smallest bias and rmse and the most informative confidence interval.

Naturally no set of simulations will be dispositive for the relative merits of the various eiv estimators. With this caveat in mind, I tentatively conclude that the simulations make a case for trying DetS and DetMCD in situations where measurement error is suspected, especially in cross-section data sets where it is reasonable to assume that serious eiv contamination affects a minority of the observations.

**5. A real data set**

Table 8 presents OLS and DetMCD results from the data set UN (Fox and Weisberg 2015), which reports the infant mortality rate and per-capita gross domestic product (gdp, in U. S. dollars) for 193 nations in 1998. The slope coefficients from a log-linear fit show that infant mortality (the dependent variable) varies inversely to gdp; but the DetMCD estimate is larger in magnitude than its OLS counterpart. Does this reflect eiv attenuation ? It seems likely that gdp is significantly mismeasured for a subset of countries with large underground sectors, poorly-funded data collection programs and unrealistic exchange rates vis-a-vis the dollar.

The standard errors (shown below their respective slope coefficients) are the usual Gauss-Markov estimates when the disturbances $u_{yi}$ are iid (e. g., Greene 2003, pp. 48-49). In addition, a bootstrap standard error is reported for the DetMCD coefficient and for the difference between OLS and DetMCD, which is statistically significant. Proposals for robust-regression bootstraps have been developed by Salibian-Barrera and Zamar (2002) and Willems and Van Aelst (2005).

## 6. Caveats and Conclusions

Eiv is a long-standing, difficult issue in linear regression; and progress depends in part on new identifying assumptions. In this spirit, I characterize measurement error as bad-leverage points and assume that fewer than half the sample observations are heavily contaminated, in which case a high-breakdown estimator may be able to isolate and downweight or discard the problematic data. I explored simulations of simple and multiple regression where eiv affected 25% of the data and $R^2$ was mediocre. DetS and the initial DetMCD estimates, although inefficient, had the smallest biases; and their 95-percent confidence intervals usually contained the true values of the slope coefficient(s).

In view of its canonical status and pedagogical value, I have emphasized the bivariate eiv model (1). However, Tables 4 and 5 make the point that high-breakdown estimation is especially advantageous for multiple linear regression, where scatter plots are less effective in detecting bad-leverage observations and where conventional outlier diagnostics can be quite misleading (Rousseeuw and Leroy 1987, chapter 3 and 6).

DetMCD and DetS are relatively new algorithms, and the conditions under which they are consistent estimators have not yet been established. In addition, it must be emphasized that they are designed to analyze data sets of *continuous-valued variables*. Regressors in the form of binary (dummy) variables pose computational and conceptual problems for high-breakdown estimation. The computational problem can be addressed if a Huber-type M estimator is used to "partial out" the dummies from each continuous-valued variable before DetMCD or DetS is applied (Maronna et al. 2006, chapter 4 and pp. 361-362). The conceptual problem has been discussed by Mili and Coakley (1996) and Hubert (1997). Dummy variables partition a data set; and although contamination (including measurement error) may affect only a fraction of the sample, an estimator can break down if the bad observations are concentrated in one or a few partitions.

## Appendix: heteroscedasticity as vertical outliers

Model (1) assumes that the disturbances $u_{yi}$ are homoscedastic; they all have the same variance: $\sigma_i^2 = \sigma^2$. If on the other hand their variances differ, the disturbances are said to be heteroscedastic. Then OLS is inefficient, and the usual formulas for the coefficients' standard errors are incorrect. Weighted least squares would be efficient if each $\sigma_i^2$ were known, but this is seldom the case in practice. Instead it is usually assumed that heteroscedasticity may affect all the observations and that $\sigma_i^2$ is correlated with one or several regressors. Consistent estimators of the standard errors are proposed, and these procedures are said to be "robust" insofar as a researcher need not specify in detail the covariance structure that links $\sigma_i^2$ to the regressors (e. g., Greene 2003, chapters 10 and 11; Davidson and MacKinnon 1993, chapter 16). However, the estimators are certainly not robust to anomalous observations since they are based on the squared OLS residuals and sometimes also on the OLS projection matrix (the "hat" matrix), both of which have very low breakdown points (Rousseeuw and Leroy 1987, pp. 216-227).

Just as measurement errors are practically indistinguishable from bad leverage points, a large $\sigma_i^2$ can be associated with a vertical outlier, defined in section 3 as an atypical observation on the dependent variable. While a heteroscedastic pattern is sometimes evident –for example, when the variability of household expenditures increases with income— vertical outliers often affect a minority of observations; and the pattern, if it exists, may be difficult to discern. A high-breakdown estimator of linear regression can be applied to these data sets.

Portnoy and Welsh (1992) show that heteroscedasticity not only distorts estimates of standard errors but may also produce estimated coefficients that are biased and inconsistent. This can occur if (i) $\sigma_i^2$ varies with a regressor, (ii) $u_{yi}$ has a skewed distribution, and (iii) $E(u_{yi}) \neq 0$. In one of the authors' bivariate models, $u_{yi}$ has an exponential distribution whose median is zero but whose expected value is positive. The regression is model (1) with $\alpha = 0$, $\beta = 1$, and heteroscedastic disturbances $\exp(-x_i)u_{yi}$. In my simulation of their model, $x_i \sim N(0,2)$ and the heteroscedasticity is limited to 25 percent of the sample. One thousand replications with n = 2000 produced an average OLS slope estimate of 0.45 with a standard error of 0.95. The heteroscedasticity-consistent standard error (HC3) was 0.50 (Zeileis 2017). In the absence of heteroscedasticity, this regression would have a rather high R-squared, 0.80. Accordingly, I chose MM as the robust estimator. The average MM slope estimate was 1.00 with a standard error of 0.01.

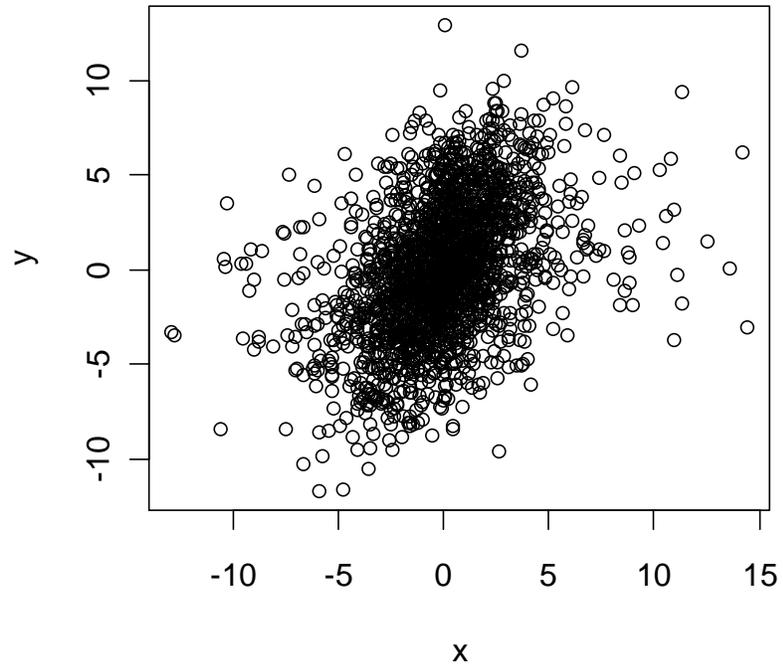

Table 1. Three high-breakdown estimators

| Estimator | Reference | R version |
|---|---|---|
| MM-estimate(MM) | Maronna et al. 2006, chapter 5 | package 'robust' Konis 2017* |
| Deterministic minimum covariance determinant (DetMCD) | Hubert et al. 2012 | package DetMCD Vakili 2018 |
| Deterministic S-type estimator (DetS) | Hubert et al. 2015 | package 'rrcov' Todorov 2018 |

*also package 'robustbase' (Maechler 2018)

Table 2. Bivariate regression with true $R^2 = 0.80$

|  | n = 200 | | | n = 2000 | | |
|---|---|---|---|---|---|---|
|  | bias | rmse | 95% c.i. | bias | rmse | 95% c.i. |
| OLS | -0.85 | 0.85 | 0.10,0.20 | -0.85 | 0.85 | 0.13,0.17 |
| MM | -0.00 | 0.06 | 0.89,1.11 | -0.01 | 0.04 | 0.96,1.03 |
| DetMCD | -0.03 | 0.13 | 0.74,1.22 | -0.01 | 0.04 | 0.91,1.08 |
| DetS | -0.01 | 0.05 | 0.89,1.09 | -0.01 | 0.02 | 0.96,1.03 |

note: $u_{xi} \sim N(10,4)$   $u_{yi} \sim N(0,1)$   $x_i \sim N(0,2)$

Table 3. Bivariate regression with true $R^2 = 0.31$

|  | n = 200 | | | n = 2000 | | |
|---|---|---|---|---|---|---|
|  | bias | rmse | 95% c.i. | bias | rmse | 95% c.i. |
| OLS | -0.85 | 0.85 | 0.06,0.24 | -0.85 | 0.85 | 0.12,0.18 |
| MM | -0.85 | 0.85 | -0.04,0.37 | -0.85 | 0.85 | 0.09,0.22 |
| DetMCD | -0.13 | 0.38 | 0.16,1.55 | -0.04 | 0.13 | 0.73,1.22 |
| DetS | -0.07 | 0.17 | 0.61,1.23 | -0.07 | 0.09 | 0.84,1.03 |

note: $u_{xi} \sim N(10,4)$   $u_{yi} \sim N(0,3)$   $x_i \sim N(0,2)$

Table 4. Multiple regression with true $R^2$ = 0.35, n = 1000

|        | x coefficient | | | z coefficient | | |
|--------|------|------|-----------|------|------|-----------|
|        | bias | rmse | 95% c.i.  | bias | rmse | 95% c.i.  |
| OLS    | -0.54 | 0.54 | 0.36,0.57 | 0.27 | 0.29 | 1.09,1.45 |
| MM     | -0.49 | 0.50 | 0.26,0.79 | 0.25 | 0.31 | 0.91,1.61 |
| DetMCD | -0.15 | 0.26 | 0.43,1.26 | 0.08 | 0.31 | 0.46,1.65 |
| DetS   | -0.23 | 0.25 | 0.60,0.92 | 0.12 | 0.17 | 0.89,1.36 |

note: $u_{xi} \sim N(0,4)$   $u_{yi} \sim N(0,4)$   $x_i, z_i \sim N(0,2)$   $r_{xz} = 0.40$

Table 5. Multiple regression with true $R^2 = 0.35$, n = 1000

|  | x coefficient | | | z coefficient | | |
|---|---|---|---|---|---|---|
|  | bias | rmse | 95% c.i. | bias | rmse | 95% c.i. |
| OLS | -0.29 | 0.29 | 0.60,0.84 | -0.22 | 0.24 | 0.61,0.94 |
| MM | -0.18 | 0.25 | 0.51,1.15 | -0.20 | 0.28 | 0.43,1.20 |
| DetMCD | -0.05 | 0.24 | 0.50,1.41 | -0.03 | 0.30 | 0.40,1.54 |
| DetS | -0.08 | 0.12 | 0.75,1.10 | 0.07 | 0.13 | 0.72,1.15 |

note: $u_{xi} \sim N(0,4)$   $u_{zi} \sim N(0,3)$   $u_{yi} \sim N(0,4)$   $x_i, z_i \sim N(0,2)$   $r_{xz} = 0.40$

Table 6. Bivariate regression with true $R^2 = 0.31$

n = 1000

|  | bias | rmse | 95% c.i. |
|---|---|---|---|
| OLS | -0.36 | 0.36 | 0.55,0.73 |
| OR | 1.15 | 1.17 | 1.91,2.48 |
| GEOM | 0.44 | 0.45 | 1.36,1.53 |
| DetMCD | -0.12 | 0.23 | 0.50,1.31 |

note:   $u_{xi} \sim N(0,3)$    $u_{yi} \sim N(0,3)$    $x_i \sim N(0,2)$

Table 7. Bivariate regression with true $R^2 = 0.31$

n = 1000

|  | bias | rmse | 95% c.i. |
|---|---|---|---|
| OLS | -0.33 | 0.34 | 0.59,0.74 |
| IV | -0.22 | 0.25 | 0.52,1.01 |
| DetMCD | -0.11 | 0.22 | 0.52,1.26 |

note:  $u_{xi} \sim N(0,4)$   $u_{yi} \sim N(0,3)$   $x_i \sim \chi^2(4\ d.f.)$

**Table 8. Infant mortality and gdp**
(n = 193)

|  | gdp | R-squared |
|---|---|---|
| OLS | -0.493 | 0.656 |
|  | 0.026 |  |
|  |  |  |
| DetMCD | -0.733 |  |
|  | 0.021 |  |
|  | 0.018 | bootstrap |
|  |  |  |
| DetMCD - OLS | -0.240 |  |
|  | 0.016 | bootstrap |